\documentclass[twocolumn,floatfix,prb,superscriptaddress]{revtex4}%
\usepackage{graphicx,graphics,color,epsfig}
\usepackage{bm}
\usepackage{amsmath}
\usepackage{mathrsfs}
\usepackage{amssymb}
\usepackage{subfigure}
\usepackage{amsfonts}
\usepackage{graphicx}
\usepackage{multirow,booktabs}
\usepackage{epstopdf}

\newcommand{\be}{\begin{eqnarray}}
\newcommand{\ee}{\end{eqnarray}}

\providecommand{\U}[1]{\protect\rule{.1in}{.1in}}
\begin{document}
\title{Quaternary Jordan-Wigner mapping and topological extended-kink phase in the interacting Kitaev chain with geometrical ring frustration}
\author{Zhen-Yu Zheng}
\affiliation{College of Physics, Sichuan University, 610064, Chengdu, People’s Republic of China\\
and Key Laboratory of High Energy Density Physics and Technology of Ministry of Education, Sichuan University, 610064,
Chengdu, People’s Republic of China}
\author{Han-Chuan Kou}
\affiliation{College of Physics, Sichuan University, 610064, Chengdu, People’s Republic of China\\
and Key Laboratory of High Energy Density Physics and Technology of Ministry of Education, Sichuan University, 610064,
Chengdu, People’s Republic of China}
\author{Peng Li}
\email{lipeng@scu.edu.cn}
\affiliation{College of Physics, Sichuan University, 610064, Chengdu, People’s Republic of China\\
and Key Laboratory of High Energy Density Physics and Technology of Ministry of Education, Sichuan University, 610064,
Chengdu, People’s Republic of China}

\date{\today}

\begin{abstract}
  On a ring, a single Jordan-Wigner transformation between the Kitaev model and the spin model suffers redundant degrees of freedom. However, we can establish an exact quaternary Jordan-Wigner mapping involving two Kitaev rings and two spin rings with periodic or antiperiodic boundary conditions. This mapping facilitates us to demonstrate exactly how a topological extended-kink (TEK) phase develops in the interacting Kitaev ring with odd number of lattice sites.  The emergence of this new phase is attributed to the effect of geometrical ring frustration. Unlike the usual topological phases protected by energy gap in noninteracting systems, the TEK phase is gapless. And because the spectra of low energy excitations are quadratic, the specific heat per site approaches a half of Boltzmann constant near absolute zero temperature. More interestingly, the ground state is unique, immune to spontaneous symmetry breaking. It exhibits a long-range correlation function with a nonlocal factor, but no local order parameter can be defined. As a concomitant effect, a special kind of localized kink zero mode (KZM) takes place if we introduce a type of bond defect. We also show that the KZM is robust against moderate disorders.
\end{abstract}

%\pacs{05.50.+q, 75.50.Ee, 02.30.Tb}
\maketitle
%75.10.Jm  Quantized spin models, including quantum spin frustration
%05.50.+q  Lattice theory and statistics (Ising, Potts, etc.)
%03.65.Ud  Entanglement and quantum nonlocality (e.g. EPR paradox, Bell's inequalities, GHZ states, etc.)
%75.50.Ee  Antiferromagnetics

%%%%%%%%%%%% FIG 1  %%%%%%%%%%%%%%%%%%%%%%%%%%%%%%%
%\begin{figure}[t]
%\begin{center}
%\includegraphics[width=1.8in,angle=0]{figure1.eps}
%\end{center}
%\caption{(Color online) A periodic spin chain. Ring frustration occurs if the number of spins is odd and the nearest-neighbor interactions %are antiferromagnetic.}%
%\label{fig1}%
%\end{figure}
%%%%%%%%%%%%%%%%%%%%%%%%%%%%%%%%%%%%%%%%%%%%%%
%\begin{widetext}

\section{Introduction}

Interacting Kitaev chain attracts much attention recently \cite{Kitaev, Turner, Rahmani, Katsura2015, Katsura2018}. It was pointed out that the classification of topological phases in noninteracting fermionic systems \cite{Schnyder, Kitaev2009} may not apply to the interacting ones \cite{Kitaev}. At a symmetric point, Miao \emph{et al.} found the problem can be solved exactly by a scheme of two-step Jordan-Wigner transformations (JWT's) \cite{Miao}. And plentiful phases have been uncovered in the dimerized case \cite{Ezawa, Chen2017, Chitov}. On the other hand, geometrical frustration first introduced in the classical Ising spin systems plays an important role in quantum antiferromagnetism \cite{Diep}. Usually the geometrical frustration does not outstand in a pure fermionic system, although similar concepts of frustration can be introduced in some fermionic systems \cite{Katsura2015, Katsura2018, Nie}.

In this work, we demonstrate that a so-called geometrical ring frustration (GRF) can play a vital role in the interacting fermionic Kitaev chain with a closed boundary condition \cite{Bariev, Cabrera, Campostrini, DongJSM, DongPRE, Franchini, He2017, Li2019}. A novel topologically nontrivial phase and concomitant zero modes is uncovered as an effect of GRF. One of our motivations comes from the fact that these kind of models may be mimicked on manmade finite systems realized by the state-of-the-art experimental techniques with flexible control methods \cite{Kwon, Labuhn, Lienhard}.

The contents are arranged as follows. In Sec. II, we demonstrate that a quaternary Jordan-Wigner mapping holds among our aimed interacting Kitaev ring and its relevant systems. In Sec. III, we illustrate a novel topological extended-kink (TEK) phase by a clear perturbative treatment first, then solve the interacting Kitaev chain at its symmetric points to exactly demonstrate its existence. In Sec. IV, we list several intriguing properties of the TEK phase, such as gaplessness of low-energy excitations, the ground state's correlation function and entanglement entropy, and the low-temperature specific heat. In Sec. V, a kind of kink zero mode (KZM) is introduced as a concomitant effect of the GRF. The robustness of the KZM against moderate disorders is also discussed. At last, we give a summary in Sec. VI.

\section{The model}

\subsection{The aimed Hamiltonian}

We aim at the interacting Kitaev ring,
\be
  H_{\mathrm{R}}(c)=\sum_{j=1}^{N}[(-t c_{j}^{\dag}c_{j+1}+\Delta c_{j}c_{j+1}+\mathrm{h.c.})\nonumber\\
  +U (2 n_{j}-1)(2 n_{j+1}-1)],  \label{HRc}
\ee
where $c=\{c_{j},(j=1,2,...,N)\}$ denotes all fermions residing on the lattice sites and $n_{j}=c_{j}^{\dag}c_{j}$. The subscript R means the periodic boundary condition (PBC),  $c_{N+1}=c_{1}$, i.e. the "Ramond" sector \cite{FrancescoBook}. In Eq. (\ref{HRc}), $U>0$ and $N\in \mathrm{Odd}$ is demanded, which ensures the effect of GRF. We note that this peculiar Hamiltonian possesses the translational symmetry, but does not the particle-hole symmetry.

\subsection{Relevant Hamiltonians and quaternary Jordan-Wigner mapping}

%%%%%%%%%%% FIG 1  %%%%%%%%%%%%%%%%%%%%%%%%%%%%%%%
\begin{figure}[t]
  \begin{center}
  \includegraphics[width=2in,angle=0]{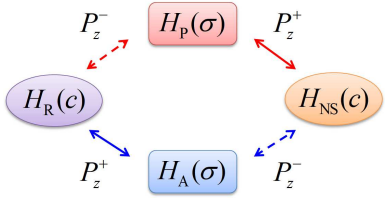}
  \end{center}
  \caption{(Color online) The quaternary Jordan-Wigner mapping among four relevant Hamiltonians, as expressed by Eqs. (\ref{mappingR}), (\ref{mappingNS})-(\ref{mappingA}).  Each arrow represents a valid parity channel.}%
  \label{fig1}%
\end{figure}
%%%%%%%%%%%%%%%%%%%%%%%%%%%%%%%%%%%%%%%%%%%%%%

First of all, one should keep in mind that the PBC case is totally different from that with open boundary condition (OBC), although the Jordan-Wigner transformation (JWT) can be applied to both of them. Because a boundary term arises in the PBC case, which imposes a strong parity constraint \cite{Lieb}. And by considering a chain with GRF, we have shown that the difference becomes even more prominent \cite{DongJSM}. In short, for the OBC case, the fermionic Hamiltonian can be mapped to a spin $XYZ$ chain exactly and vice versa \cite{Lieb}. While for the PBC case we are focusing here, the transformation suffers a problem of redundant degrees of freedom (DOF). To fix this problem and to solve the interacting Kitaev ring, we introduce a complete \emph{quaternary Jordan-Wigner mapping} (QJWM).

Let us see the application of JWT,
\be
  &c_{j}^{\dag}=\frac{1}{2}(\sigma_{j}^{x}+i \sigma_{j}^{y})\prod_{l=1}^{j-1}(-\sigma_{l}^{z}), (1\leq j\leq N), \label{JWT}
\ee
in detail. It leads to the spin $XYZ$ ring with an extraordinary boundary term \cite{Lieb},
\be
  H_{R}(c)\rightarrow H+H_{N},
\ee
where
\be
  &H=\sum_{j=1}^{N-1}\left(-\frac{t+\Delta}{2} \sigma_{j}^{x} \sigma_{j+1}^{x}\right.  -\frac{t-\Delta}{2} \sigma_{j}^{y} \sigma_{j+1}^{y}+U \sigma_{j}^{z} \sigma_{j+1}^{z} ), \\
  &H_{N}=\mathscr{P}_{z} (-\frac{t+\Delta }{2}\sigma_{N}^{x}\sigma_{1}^{x}-\frac{t-\Delta }{2}\sigma_{N}^{y}\sigma_{1}^{y}) +U\sigma_{N}^{z}\sigma_{1}^{z},
\ee
with
\be
  \mathscr{P}_{z} &=& \exp(i \pi M_{z}),\\
  M_{z} &=& \sum_{j=1}^{N}c_{j}^{\dag} c_{j}=\sum_{j=1}^{N}\frac{1+\sigma_{j}^{z}}{2}.
\ee
The parity operator $\mathscr{P}_{z}$ imposes a strong constraint on the mapping between the fermionic Hilbert space and the spin Hilbert space. It means that the solution of $H_{R}(c)$ involves two $XYZ$ rings denoted by
\be
  &H_{\mathrm{P}}(\sigma)=H+ (-\frac{t+\Delta }{2}\sigma_{N}^{x}\sigma_{1}^{x}-\frac{t-\Delta }{2}\sigma_{N}^{y}\sigma_{1}^{y}) +U\sigma_{N}^{z}\sigma_{1}^{z},\\
  &H_{\mathrm{A}}(\sigma)=H- (-\frac{t+\Delta }{2}\sigma_{N}^{x}\sigma_{1}^{x}-\frac{t-\Delta }{2}\sigma_{N}^{y}\sigma_{1}^{y}) +U\sigma_{N}^{z}\sigma_{1}^{z},
\ee
where $\sigma=\{\sigma_{j}^{a},(j=1,2,...,N; a=x,y,z)\}$ denotes all Pauli spins. The subscript P means PBC in transverse directions, $\sigma_{N+1}^{a}=\sigma_{1}^{a}$ ($a=x,y$), while the subscript A means anti-PBC in the transverse directions $\sigma_{N+1}^{a}=-\sigma_{1}^{a}$ ($a=x,y$). Please notice that both cases exhibit PBC in the longitudinal direction, $\sigma_{N+1}^{z}=\sigma_{1}^{z}$. The total DOF of $H_{\mathrm{P}}(\sigma)$ and $H_{\mathrm{A}}(\sigma)$ are twice as that of $H_{\mathrm{R}}(c)$. To eliminate the redundant DOF, we can apply the projection,
\be
  &H_{\mathrm{R}}(c)=P_{z}^{-}H_{\mathrm{P}}(\sigma)P_{z}^{-}+P_{z}^{+}H_{\mathrm{A}}(\sigma)P_{z}^{+}, \label{mappingR}
\ee
where the projectors are defined as
\be
  P_{z}^{\pm}=\frac{1}{2}(1\pm\mathscr{P}_{z}).
\ee
Such a procedure has been noticed by many researchers in previous studies \cite{Lieb, Suzuki}.

Now we may ask where do the redundant DOF go? If defining a new fermionic Hamiltonian,
\be
  H_{\mathrm{NS}}(c)&=&\sum_{j=1}^{N-1}(-t c_{j}^{\dag}c_{j+1}+\Delta c_{j}c_{j+1}+\mathrm{h.c.})\nonumber\\
  &-&(-t c_{N}^{\dag}c_{1}+\Delta c_{N}c_{1}+\mathrm{h.c.})\nonumber\\
  &+&\sum_{j=1}^{N}U (2 n_{j}-1)(2 n_{j+1}-1),  \label{HNSc}
\ee
where the subscript NS means the anti-PBC,  $c_{N+1}=-c_{1}$, i.e. the "Neveu-Schwarz" sector \cite{FrancescoBook}, then one finds it is easy to verify another projection,
\be
  &H_{\mathrm{NS}}(c)=P_{z}^{+}H_{\mathrm{P}}(\sigma)P_{z}^{+}+P_{z}^{-}H_{\mathrm{A}}(\sigma)P_{z}^{-}. \label{mappingNS}
\ee
We can also obtain the inverse projections of Eqs. (\ref{mappingR}) and (\ref{mappingNS}),
\be
  &H_{\mathrm{P}}(\sigma)=P_{z}^{-}H_{\mathrm{R}}(c)P_{z}^{-}+P_{z}^{+}H_{\mathrm{NS}}(c)P_{z}^{+}, \label{mappingP}\\
  &H_{\mathrm{A}}(\sigma)=P_{z}^{+}H_{\mathrm{R}}(c)P_{z}^{+}+P_{z}^{-}H_{\mathrm{NS}}(c)P_{z}^{-}. \label{mappingA}
\ee
Thus the four involved Hamiltonians, $H_{\mathrm{R/NS}}(c)$ and $H_{\mathrm{P/A}}(\sigma)$, satisfy a quaternary mapping as illustrated in Fig. 1. The full mapping runs out of all DOF of the four Hamiltonians without any redundancy. We also notice that the two fermionic Kitaev rings can be transformed to each other, $\mathscr{C}^{\dag} H_{\mathrm{R}}(c) \mathscr{C} = H_{\mathrm{NS}}(c)$, by the particle-hole conjugation operator $\mathscr{C}=\prod_{j=1}^{N}[c_{j}^{\dag}+(-1)^{j}c_{j}]$. Both $H_{\mathrm{R}}(c)$ and $H_{\mathrm{NS}}(c)$ do not possess particle-hole symmetry. This fact is in contrast to the case in OBC problem \cite{Miao}.

\section{Emergence of the TEK phase}

The QJWM plays an important role in disclosing the phases in $H_{\mathrm{R}}(c)$. In this section, we first provide a clear picture for the newly discovered TEK phase by a perturbative treatment, then solve $H_{\mathrm{R}}(c)$ at its symmetric points to verify the TEK phase exactly.

\subsection{The effect of GRF: perspective from a perturbative treatment}

For large $U$, we may take the interacting term as the dominant part of $H_{\mathrm{R}}(c)$,
\be
  H_{0}(c)=\sum_{j=1}^{N}U (2 n_{j}-1)(2 n_{j+1}-1). \label{H0c}
\ee
It is equivalent to the simple classical antiferromagnetic Ising chain with GRF. Its solutions can be categorized as one-kink states, three-kink states, etc. The $2N$ one-kink states serve as the degenerate ground states. There is a big gap of magnitude $4U$ to the upper energy level. Noticing that the $c$ fermion number representation and the spin $\sigma^z$ representation are identical, i.e. $|0_{j}\rangle=|\downarrow_{j}\rangle$ and $|1_{j}\rangle=|\uparrow_{j}\rangle$, we denote the $2N$ one-kink states conveniently as
\be
  &|j,\uparrow\rangle = |\cdots,\downarrow_{j-1},\boxed{\uparrow_{j},\uparrow_{j+1}},\downarrow_{j+2},\cdots\rangle, \\
  &|j,\downarrow\rangle = |\cdots,\uparrow_{j-1},\boxed{\downarrow_{j},\downarrow_{j+1}},\uparrow_{j+2},\cdots\rangle,
\ee
in which the boxes indicate where the classical kinks are. They are also eigenstates of the parity operator $\mathscr{P}_{z}$, which means that the diagonalization of the problem can be performed in the two subspaces with $\mathscr{P}_{z}=\pm1$ separately.

%%%%%%%%%%% FIG 2  %%%%%%%%%%%%%%%%%%%%%%%%%%%%%%%
\begin{figure}[t]
  \begin{center}
  \includegraphics[width=2.3in,angle=0]{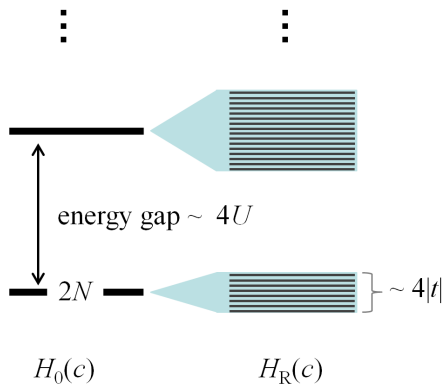}
  \end{center}
  \caption{(Color online) The schematics of the emergence of the TEK phase in the interacting Kitaev ring $H_{R}(c)$. The formation of the lowest band of $H_{R}(c)$ can be attributed to the quantum fluctuations disturbing the classical Ising ring $H_{0}(c)$ with GRF.}%
  \label{fig2}%
\end{figure}
%%%%%%%%%%%%%%%%%%%%%%%%%%%%%%%%%%%%%%%%%%%%%%

Then we may take the rest of the Hamiltonian as a perturbation, which arouses quantum fluctuations. According to the projection Eq. (\ref{mappingR}), we can conveniently perform the calculations on $P_{z}^{-}H_{\mathrm{P}}P_{z}^{-}$ and $P_{z}^{+}H_{\mathrm{A}}P_{z}^{+}$ and accomplish the projection to get the solution. For $P_{z}^{-}H_{\mathrm{P}}P_{z}^{-}$, if we cut off the Hilbert space in the subspace of one-kink states, we shall arrive at the effective Hamiltonian
\begin{align}
  P_{z}^{-}H_{\mathrm{P}}(\sigma)P_{z}^{-}\approx& P_{z}^{-}H_{\mathrm{P}}^{\mathrm{eff}}(\sigma)P_{z}^{-} \nonumber\\
=&P_{z}^{-}\sum_{j,\tau}^{N}\{(2-N)U|j,\tau\rangle\langle j,\tau|
       \nonumber\\
&- t(|j,\tau\rangle\langle j+2,\tau|+\mathrm{h.c.})\}P_{z}^{-}.
\end{align}
It is easy to check that the states,
\be
  |E_{\mathrm{P}}(q)\rangle& = \frac{1}{\sqrt{N}} \sum_{j} e^{-i q j} |j,\tau\rangle
\ee
with $\tau=\uparrow$ for $N=4K+1$ and  $\tau=\downarrow$ for $N=4K+3$, are valid ones with odd parity for the effective Hamiltonian. The energy spectrum of $|E_{\mathrm{P}}(q)\rangle$ reads
\be
  E_{\mathrm{P}}(q) = -(N-2)U-2 t \cos(2q), \label{EP(q)approx},
\ee
where
\be
  &q\in \{-\frac{N-1}{N}\pi,\ldots,-\frac{2}{N}\pi,0,\frac{2}{N}\pi,\ldots,\frac{N-1}{N}\pi\}.
\ee
Likewise, for $P_{z}^{+}H_{\mathrm{A}}P_{z}^{+}$, we can get the other $N$ valid states with even parity,
\be
  |E_{\mathrm{A}}(q)\rangle& = \frac{1}{\sqrt{N}} \sum_{j} e^{-i q j} |j,\tau\rangle
\ee
with $\tau=\downarrow$ for $N=4K+1$ and  $\tau=\uparrow$ for $N=4K+3$. The corresponding energy spectrum reads
\be
  E_{\mathrm{A}}(q) = -(N-2)U+2 t \cos(2q). \label{EA(q)approx}
\ee
In Eqs. (\ref{EP(q)approx}) and (\ref{EA(q)approx}), the higher-order contributions up to $O(t/U)^2$ and $O(\Delta/U)^2$ are neglected due to the cutoff. For any $N\in\mathrm{Odd}$, the non-degenerate ground state in the one-kink approximation reads
\be
  |E_0\rangle = \left\{\begin{array}{ll}
    |E_{\mathrm{P}}(0)\rangle, (\mathrm{for}~t>0),\\
    |E_{\mathrm{A}}(0)\rangle, (\mathrm{for}~t<0),
  \end{array}\right. \label{E0perturbation}
\ee

Thus the degeneracy of the $2N$ Ising states is lifted and a gapless band of width about $4|t|$ comes into being under the condition of thermodynamic limit, $N\rightarrow\infty$. The one-kink states prevail in this lowest band, although the states with more kinks will blend in if we include more above energy states. However, the physical picture of energy level splitting will not change (Fig. 2). The lowest quasi-continuous energy band forming by $2N$ levels can keep robust in a considerable range of model parameters, at least for $t, \Delta \ll U$.

Below, through the exact solution at a symmetric point, we will see that the extended-kink states are topologically nontrivial within the picture of fermionic Kitaev ring. So this novel phase induced mainly by the GRF and quantum fluctuations is named as \emph{topological extended-kink} (TEK) phase.

\subsection{Exact solution of $H_{\mathrm{R}}(c)$ at the \\symmetric points $t=\pm\Delta$}

%%%%%%%%%%% FIG 3  %%%%%%%%%%%%%%%%%%%%%%%%%%%%%%%
\begin{figure}[t]
  \begin{center}
  \includegraphics[width=2.2in,angle=0]{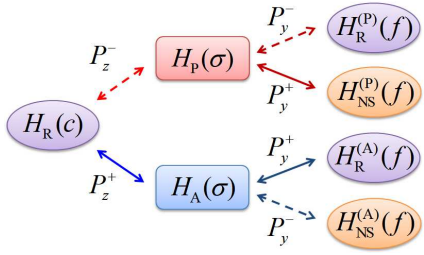}
  \end{center}
  \caption{(Color online) The scheme of two-step Jordan-Wigner transformations for solving the Kitaev ring $H_{\mathrm{R}}(c)$ at the symmetric point $t=\Delta$. Six auxiliary Hamiltonians are involved.}%
  \label{fig3}%
\end{figure}
%%%%%%%%%%%%%%%%%%%%%%%%%%%%%%%%%%%%%%%%%%%%%%

Now let us aim at the interacting Kitaev ring $H_{\mathrm{R}}(c)$ at its symmetric line $t=\Delta$. Another symmetric line $t=-\Delta$ is almost the same. We will talk about it later. In this situation, the system can be solved exactly by a scheme of two-step JWT's similar to the one for OBC problem \cite{Miao}. However, the procedure is much more cumbersome here, because six auxiliary Hamiltonians are involved. And because complicated projections are imposed, the solution does not fit for a conventional free fermion picture. According to the QJWM, the scheme is designed as shown in Fig. 3. By the first step of JWT, Eq. (\ref{JWT}), the resulting Hamiltonians, $H_{\mathrm{P}}(\sigma)$ and $H_{\mathrm{A}}(\sigma)$, are spin $XZ$ rings in fact. Then in the second step, we can perform the JWT,
\be
  f_{j}^{\dag}=\frac{1}{2}(\sigma_{j}^{z}+i \sigma_{j}^{x})\prod_{l=1}^{j-1}(-\sigma_{l}^{y}),
\ee
for $H_{\mathrm{P}}(\sigma)$ and another JWT,
\be
  f_{j}^{\dag}=\frac{1}{2}\left[\sigma_{j}^{z}+i (-1)^{j+1}\sigma_{j}^{x}\right]\prod_{l=1}^{j-1}(-\sigma_{l}^{y}),
\ee
for $H_{\mathrm{A}}(\sigma)$ respectively, so as to get four Hamiltonians,
\be
  &H_{\mathrm{s}}^{(\mathrm{P})}(f)=H(\mathrm{s},-t+U,t+U), \label{Hr1f}\\
  &H_{\mathrm{s}}^{(\mathrm{A})}(f)=H(\mathrm{s},t+U,-t+U), \label{Hr2f}
\ee
where $\mathrm{s}$ refers to $\mathrm{R}$ or $\mathrm{NS}$. Both Eqs. (\ref{Hr1f}) and (\ref{Hr2f}) fall into the same general non-interacting Hamiltonian \cite{Kitaev2001, Katsura-TBC},
\begin{align}
  H(\mathrm{s},a,b)=\sum_{q\in Q_{\mathrm{s}}, q\neq q_{\mathrm{s}}}\phi_{q}^{\dag}
  (h_{y} \sigma^{y}+h_{z} \sigma^{z})\phi_{q}  \nonumber\\
  +a \cos q_{\mathrm{s}} (2f_{q_{\mathrm{s}}}^{\dag}f_{q_{\mathrm{s}}}-1), \label{HsD}
\end{align}
where $\sigma^{y}$ and $\sigma^{z}$ are Pauli matrices,
\be
  &\phi_{q}^{\dag}=(f_{q}^{\dag}, f_{-q}),\\
  &h_y=-b \sin q, h_z=-a\cos q,\\
  &q_{\mathrm{R}}=0, q_{\mathrm{NS}}=\pi,\\
  &Q_{\mathrm{R}}=\{-\frac{N-1}{N}\pi,\ldots,-\frac{2}{N}\pi,0,\frac{2}{N}\pi,\ldots,\frac{N-1}{N}\pi\}, \\
  &Q_{\mathrm{NS}}=\{-\frac{N-2}{N}\pi,\ldots,-\frac{1}{N}\pi,\frac{1}{N}\pi,\ldots,\frac{N-2}{N}\pi,\pi\}.
\ee
The associated projectors are
\be
  P_{y}^{\pm}=\frac{1}{2}(1\pm\mathscr{P}_{y}),
\ee
where
\be
  \mathscr{P}_{y} &=& \exp(i \pi M_{y}),\\
  M_{y}&=&\sum_{j=1}^{N} f_{j}^{\dagger}f_{j}=\sum_{j=1}^{N}\frac{1+\sigma_{j}^{y}}{2}.
\ee
At last, the solution of $H_{\mathrm{R}}(c)$ is retrieved by filtering out redundant DOF of the auxiliary Hamiltonians following the scheme in Fig. 3 (Please see details in Appendix A).

%%%%%%%%%%% FIG 4  %%%%%%%%%%%%%%%%%%%%%%%%%%%%%%%
\begin{figure}[b]
  \begin{center}
  \includegraphics[width=3.3in,angle=0]{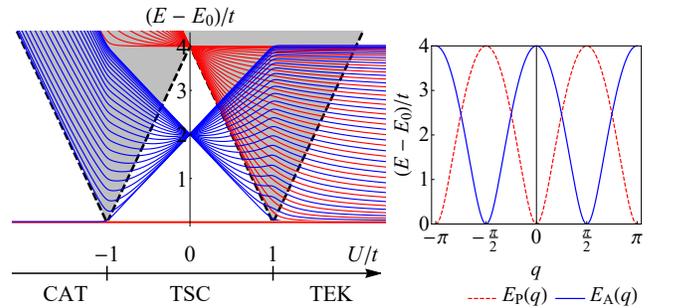}
  \end{center}
  \caption{(Color online) \emph{Left}: Low-lying energy levels and ground-state phase diagram of $H_{\mathrm{R}}(c)$ along the symmetric line $t=\Delta$. We choose $N=51$ for demonstration. The red lines represent levels of odd parity, while the blue ones even parity. All other levels distribute in the shaded area above the dashed line. There are three phases: the Shr\"{o}dinger-cat-like (CAT), the topological superconducting (TSC), and the topological extended-kink (TEK) phases. Critical points occur at $U=\pm t$. \emph{Right}: The exact lowest two energy spectra in the TEK phase at $U/t=2$.}
  \label{fig4}%
\end{figure}
%%%%%%%%%%%%%%%%%%%%%%%%%%%%%%%%%%%%%%%%%%%%%%

In the parameter range $t>0$ and $-\infty<U<\infty$, the ground state is of odd parity and reads
\be
  |E_0\rangle = \frac{1}{\sqrt{2}}(f_{0}^{\dag}|\phi_{\mathrm{R}}^{(\mathrm{P})}\rangle-|\phi_{\mathrm{NS}}^{(\mathrm{P})}\rangle),
  \label{E0exact}
\ee
where $|\phi_{\mathrm{R}}^{(\mathrm{P})}\rangle$ and $|\phi_{\mathrm{NS}}^{(\mathrm{P})}\rangle$ are BCS-type vacua of $H_{\mathrm{R}}^{(\mathrm{P})}(f)$ and $H_{\mathrm{NS}}^{(\mathrm{P})}(f)$ respectively. In the thermodynamic limit, $N\rightarrow\infty$, there are two phase transition points occurring at $U=\pm t$. Three phases emerge (Fig. 4): the Schr\"{o}dinger-cat-like (CAT), topological superconductor (TSC), and topological extended-kink (TEK) states. In the CAT phase, another energy level with even parity approaches the ground state rapidly with $N$ increasing, so the ground state becomes doubly degenerate. In the TSC and TEK phases, the ground state remains unique. The CAT and TSC are gapped, while the TEK phase becomes gapless surprisingly. $2N$ low-lying energy levels take charge in the gapless TEK phase. We label them as $\{|E_{\mathrm{P/A}}(q)\rangle, \forall q\in Q_{\mathrm{R}}\}$, since they come from the two channels, $H_{\mathrm{P/A}}(\sigma)$, respectively. Now the ground state read $|E_{0}\rangle=|E_{\mathrm{P}}(0)\rangle$ for $t>0$. As shown in Fig. 4, these low energy excitations form two interweaving true spectra,
\be
  E_{\mathrm{P}}(q)&=&2\omega_{\mathrm{P}}(q)+\mathscr{E}_{0}-(U-t),\label{EP(q)}\\
  E_{\mathrm{A}}(q)&=&2\omega_{\mathrm{A}}(q)+\mathscr{E}_{0}+|U+t|-2U,\label{EA(q)}
\ee
where
\be
  \omega_{\mathrm{P}}(q)&=&\sqrt{U^2+t^2-2 U t\cos(2q)},\\
  \omega_{\mathrm{A}}(q)&=&\sqrt{U^2+t^2+2 U t\cos(2q)},\\
  \mathscr{E}_{0}&=&(U-t)-\sum_{q\in Q_{\mathrm{R}}}\omega_{\mathrm{P}}(q).
\ee
The exact results are in excellent agreement with the one-kink approximations, Eqs. (\ref{EP(q)approx}) and (\ref{EA(q)approx}), by perturbative treatment. It is easy to show that the latter can be obtained by setting $t/U\rightarrow 0$ in the former.

The nontrivial topology of the TEK phase is embodied in the four auxiliary Hamiltonians, $H_{\mathrm{R/NS}}^{(\mathrm{P})/(\mathrm{A})}(f)$, whose parameters locate in the topologically nontrivial region \cite{Kitaev2001}. Alternatively, we can work out the winding number as a topological index \cite{Chen2018} for $H_{\mathrm{R/NS}}^{(\mathrm{P})}(f)$,
\be
  w^{(\mathrm{P})}&=&\int_{0}^{2\pi} \frac{d q}{2\pi}\frac{h_{y}\partial_{q}h_{z}-h_{z}\partial_{q}h_{y}}{h_{y}^{2}+h_{z}^{2}}\nonumber\\
  &=&\left\{\begin{array}{ll}
    {-1,} & {(U<-t)}, \\
    {1,} & {(|U|<t)}, \\
    {-1,} & {(U>t)}.
  \end{array}\right.
\ee
And for $H_{\mathrm{R/NS}}^{(\mathrm{A})}(f)$, we obtain $w^{(\mathrm{A})} =-w^{(\mathrm{P})}$. From the values of the winding number, it seems all phases are topologically nontrivial. But this is not necessarily the case, because the CAT ($U<-t$) phase will become trivial due to spontaneous symmetry breaking \cite{Miao}. While the TSC ($|U|<t$) and TEK ($U>t$) phases maintain nontrivial topology since the ground state is nondegenerate and immune to symmetry breaking. It is noteworthy that, unlike the OBC case, the Majorana fermions coming from adjacent sites are paired here \cite{Kitaev2001}.

%%%%%%%%%%% FIG 5  %%%%%%%%%%%%%%%%%%%%%%%%%%%%%%%
\begin{figure}[b]
  \begin{center}
    \includegraphics[width=3.0in,angle=0]{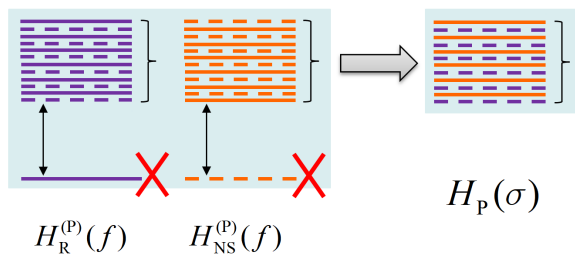}
  \end{center}
  \caption{(Color online) An alternative schematics of the emergence of the gapless TEK phase, which is different from that in Fig. 2. The lowest gapless spectra $H_{\mathrm{P}}(\sigma)$ is constructed by retaining the energy states with valid parity in $H_{\mathrm{R}}^{(\mathrm{P})}(f)$ and $H_{\mathrm{NS}}^{(\mathrm{P})}(f)$, whose ground states are discarded in the projection.}
\label{fig5}%
\end{figure}
%%%%%%%%%%% FIG 5  %%%%%%%%%%%%%%%%%%%%%%%%%%%%%%%

At this moment, one may ask why the gapless TEK phase can emerge in the interacting Kitaev ring $H_{\mathrm{R}}(c)$, while the four auxiliary noninteracting Kitaev rings $H_{\mathrm{R/NS}}^{(\mathrm{P})/(\mathrm{A})}(f)$ are gapped. We show the answer by displaying an alternative schematics for the emergence of the TEK phase in Fig. 5. When we reconstruct the true energy levels of the original problem, the ground states of $H_{\mathrm{R/NS}}^{(\mathrm{P})/(\mathrm{A})}(f)$ should be discarded. Meanwhile the valid states in the upper band are kept, so the gapless TEK phase comes into being. Fig. 5 is for the case of spin Hamiltonian $H_{\mathrm{P}}(\sigma)$, while the case of $H_{\mathrm{A}}(\sigma)$ is similar. The TEK phase of $H_{\mathrm{R}}(c)$ is obtained by the two results in a further projection (Fig. 3).

\subsection{Extended ground phase diagram for the TEK phases}

%%%%%%%%%%% FIG 6  %%%%%%%%%%%%%%%%%%%%%%%%%%%%%%%
\begin{figure}[t]
\begin{center}
\includegraphics[width=2.1in,angle=0]{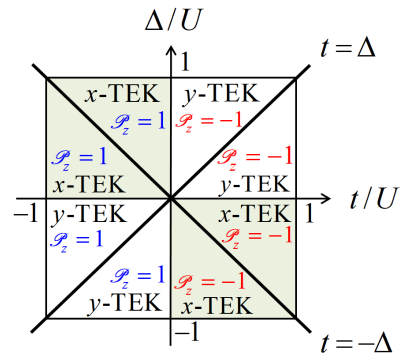}
\end{center}
\caption{(Color online) Extended ground-state phase diagram for the TEK phases. Please see details in the text.}
\label{fig6}%
\end{figure}
%%%%%%%%%%% FIG 6  %%%%%%%%%%%%%%%%%%%%%%%%%%%%%%%

Besides the exactly solvable symmetric point $t=\Delta$, there is another one at $t=-\Delta$. The former is linked to spin $XZ$ rings, the latter spin $YZ$ rings. To distinguish the associate TEK phases, we call them "$y$-TEK" and "$x$-TEK" phases respectively. For parameters apart from the symmetric points, $t\neq\pm\Delta$, we can resort to the perturbative treatment so as to expand the phase diagram for the TEK phases in the $(t/U, \Delta/U)$-parameter plane ($U>0$). The result is illustrated in Fig. 5. This phase diagram can also be verified by exact diagonalization on small lattices of odd total number of lattice sites, say $N=13$, which can help us to witness the sign for TEK phases: the lowest band composed of $2N$ energy levels \cite{DongJSM}. Notice the ground state is of even parity and comes from $H_{\mathrm{A}}(\sigma)$ when $t<0$ according to Eq. (\ref{E0perturbation}).

\section{Properties of the TEK phase}

The emergent TEK phase is quite different from the usual phases of fermionic systems. Now we uncover some of its ground-state properties and low-temperature thermodynamics.

\subsection{Correlation function of the ground state}

The correlation function of the ground state is defined as
\be
  C_{r,N}=\langle E_{0}|(2 n_{j}-1)(2 n_{j+r}-1)|E_{0}\rangle. \label{CrNapprox}
\ee
First let us see the approximate ground state for the TEK phase, Eq. (\ref{E0perturbation}), that is valid for $t,\Delta\ll U$. One can easily get the result,
\be
  C_{r,N}\approx(-1)^{r}(1-2\alpha),  \label{CrNapprox2}
\ee
where $\alpha=r/N$ ($0\leq\alpha<1/2$). Since $\alpha$ can measure a nonlocal distance $r=N \alpha$, we may call the term, $(1-2\alpha)$, the nonlocal factor \cite{Li2019}. The appearance of the nonlocal factor in the TEK phase can be verified rigorously as shown below.

For the exact ground state, Eq. (\ref{E0exact}), the correlation function can be expressed by the Toeplitz determinants as (Appendix B),
\be
  C_{r,N}=\frac{1}{2}\left(\det[\mathscr{D}^{\mathrm{(R)}}_{l-m}+\frac{2}{N}] +\det[\mathscr{D}^{\mathrm{(NS)}}_{l-m}]\right), \label{CrN}
\ee
where $1\leqslant l\leqslant r$, $1\leqslant m\leqslant r$, and
\begin{align}
  &\mathscr{D}^{\mathrm{(s)}}_{n}=-\frac{e^{-i q_{\mathrm{s}}(n+1)}}{N}+
\sum_{q\in Q_{\mathrm{s}}, q\neq q_{\mathrm{s}}}\frac{e^{-i q n}D(e^{i q})}{N},\\
  &D(e^{i q})=\frac{-(U-t e^{-2 i q})}{\sqrt{(U-t e^{-2 i q})(U-t e^{2 i q})}}.
\end{align}
The evaluation of the two Toeplitz determinants in Eq. (\ref{CrN}) is delicate. One should be very careful to keep the ratio $\alpha=r/N$ before taking the thermodynamic limit $N\rightarrow\infty$. With the help of a generalized Szeg\"{o}'s theorem \cite{DongJSM, DongPRE}, the asymptotic results at the symmetric point are obtained and read (Appendix B)
\be
  C(r,\alpha)&\equiv&\lim_{N\rightarrow\infty}C_{r,N}\nonumber\\
&\approx& \left\{\begin{array}{ll}
    {\sqrt{1-\frac{t^{2}}{U^{2}}},} & {(U<-t)}, \\
    {0,} & {(|U|<t)}, \\
    {(-1)^{r}\sqrt{1-\frac{t^{2}}{U^{2}}}(1-2\alpha),} & {(U>t)}.
  \end{array}\right. \label{Calpha}
\ee
The results for CAT and TSC states are like the ones in OBC problem \cite{Miao}. However, the result for the TEK state contains a nonlocal factor \cite{Li2019}, $(1-2\alpha)$, in excellent agreement with the result by one-kink approximation, Eq. (\ref{CrNapprox2}). The presence of the nonlocal factor signifies the absence of local order parameter of the CDW type although the correlation is long-range. The spontaneous symmetry breaking, occurring for the classical Ising Hamiltonian $H_{0}(c)$, won't occur any longer since the quantum fluctuations split the lowest $2N$ states into a band of width $\sim4t$. Consequently, in this peculiar situation, we ought not to define local order parameter by the square root of correlation function following the conventional way \cite{Yang}.

\subsection{Entanglement entropy of the ground state}

The non-degenerate ground state, Eq. (\ref{E0exact}), is highly entangled. Its one-kink approximation, Eq. (\ref{E0perturbation}), is reminiscent of the well-known generalized W state with somehow robust entanglement \cite{Cirac}. To reveal its entangled nature, we work out the entanglement entropy (EE) for the exact ground state at the symmetric point $t=\Delta$. The reduced density matrix is defined as
\be
  \rho_{l}=\mathrm{tr}_{N-l}|E_{0}\rangle\langle E_{0}|
\ee
and the EE as \cite{Vidal}
\be
  S_{l}=-\mathrm{tr}(\rho_{l}\log_{2}{\rho_{l}}).
\ee
The evaluation of EE is performed numerically \cite{DongJSM}. The results of $S_{(N-1)/2}$ are illustrated in Fig. 7. We see, in the TEK phase, the EE of the ground state is not divergent although the energy excitations are gapless.

%%%%%%%%%%% FIG 7  %%%%%%%%%%%%%%%%%%%%%%%%%%%%%%%
\begin{figure}[t]
  \begin{center}
    \includegraphics[width=3.3in,angle=0]{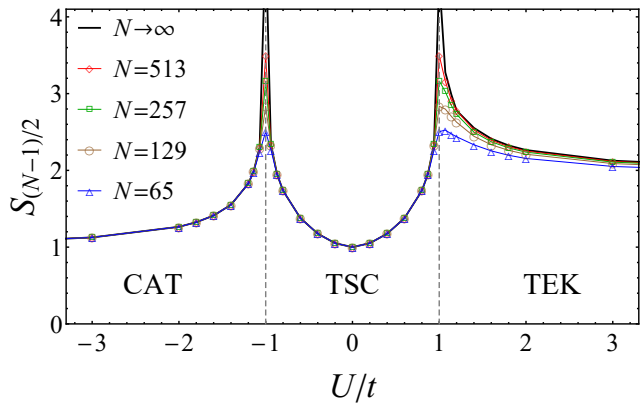}
    \caption{(Color online) Entanglement entropy (\emph{EE}), $S_{(N-1)/2}$, at the symmetric point $t=\Delta$. The black solid line is an extrapolation to the thermodynamic limit, $N\rightarrow\infty$. At the two critical points, $U/t=\pm 1$, the EE becomes divergent and obeys the CFT's prediction, $S_{(N-1)/2}^{c}\sim \frac{1}{2}+\frac{1}{3}\log_2 N$ for large enough $N$.}%
    \label{fig7}%
  \end{center}
\end{figure}
%%%%%%%%%%% FIG 7  %%%%%%%%%%%%%%%%%%%%%%%%%%%%%%%

\subsection{Density of states in the thermodynamics and specific heat at low temperatures}

The density of states (DOS) near the ground state can be worked out by Eqs. (\ref{EP(q)}) and (\ref{EA(q)}) (see also Fig. 4: \emph{Right}) exactly as
\be
  \rho(x) =\frac{2\{x+2(U-t)\}}{\pi\sqrt{x(x+4U)(4t-x)\{x+4(U-t)\}}}.
\ee
where $x=E-E_0$. It is divergent at low energies,
\be
  \rho(x) &\sim& a x^{-1/2}+b x^{1/2}+O(x^{3/2}), \\
  a&=&\frac{(U-t)^{1/2}}{2\pi (U t)^{1/2}}, \\
  b&=&\frac{U^2+U t+t^2}{16\pi (U t)^{3/2} (U-t)^{1/2}},
\ee
since the spectra are quadratic,
\be
  E_{\mathrm{P}}(q)&\sim& q^2+(q-\pi)^2,\\
  E_{\mathrm{A}}(q)&\sim& (q-\pi/2)^2+(q+\pi/2)^2.
\ee
So one can find that the specific heat per site approaches a constant when $T\ll 4t/k_{\mathrm{B}}$,
\be
  C_{M}/N=\frac{k_{B}}{2}+\frac{b k_{B}^{2}}{a} T+O(T^{2}), \label{CM}%
\ee
where $k_{\mathrm{B}}$ is the Boltzmann constant. This exotic behaviour is in contrast to the linear law in temperature, $C_M/N\sim T$, which goes to zero when $T\rightarrow 0$ no matter there are interactions or not \cite{Giamarchi}.

\section{Localized kink zero modes}

Now we display a peculiar type of zero modes within the TEK phase -- the localized \emph{kink zero modes} (KZM's), or \emph{kink bound states} \cite{explain}. First of all, we can not cut the ring, because the effect of GRF will gone. So we introduce a bond defect by altering the boundary term connecting site $N$ and $1$ in $H_{\mathrm{R}}(c)$ to
\be
  &H_{\mathrm{b}}=- t_{N}(c_{N}^{\dag}c_{1}+\mathrm{h.c.})+ \Delta_{N} (c_{N}c_{1}+\mathrm{h.c.})\nonumber\\
  &+U_{N} (2 n_{N}-1)(2 n_{1}-1). \label{HBc1}
\ee
We consider uniform interactions, $U_{N}=U$, so that the low-energy $2N$ Ising kink states are prepared and the effect of GRF is maintained. We also consider the symmetric point, $\Delta=t$ and $\Delta_{N}=t_{N}$, so that the defect is controlled by the simple ratio $\gamma=t_{N}/t$.

\subsection{Picture of the KZM's: perturbative treatment}

First, to get a clear picture, we dwell on the perturbative theory based on one-kink approximation that is reliable as proved above. We perform the perturbative treatment on the spin Hamiltonians $H_{\mathrm{P}}(\sigma)$ and $H_{\mathrm{A}}(\sigma)$ with corresponding defects. Then the solution for $H_{\mathrm{R}}(c)$ is obtained by projection according to Eq. (\ref{mappingR}). As shown in Fig. 8, we found there emerge two KZM's below the bulk states when $\gamma>1$. They come from $H_{\mathrm{P}}(\sigma)$ and $H_{\mathrm{A}}(\sigma)$ respectively and read
\be
  &|\mathrm{KZM}_{\mathrm{P}}\rangle = \sum_{j}\psi_{j}|j,\tau\rangle,\\
  &|\mathrm{KZM}_{\mathrm{A}}\rangle = \sum_{j}\chi_{j}|j,\overline{\tau}\rangle,
\ee
where
\be
  (\tau,\bar{\tau})=\left\{\begin{array}{ll}
    (\uparrow,\downarrow),~~~(\mathrm{for}~N=4K+1),\\
    (\downarrow,\uparrow),~~~(\mathrm{for}~N=4K+3),
  \end{array}\right.
\ee
to make sure that $|\mathrm{KZM}_{\mathrm{P}}\rangle$ is of odd parity and $|\mathrm{KZM}_{\mathrm{A}}\rangle$ even parity. For small $N$, $|\mathrm{KZM}_{\mathrm{A}}\rangle$ has a little higher energy than $|\mathrm{KZM}_{\mathrm{P}}\rangle$. But with $N$ increasing, they become degenerate rapidly. There is an energy gap,
\be
  \Delta_{\mathrm{g}}=(\gamma+\frac{1}{\gamma}-2)t,\label{Delta_g}
\ee
from the two KZM's to the above bulk states. For large enough $N$, the asymptotic solution reads
\begin{align}
  \psi_{j}=  \left\{\begin{array}{ll}
    {\sqrt{\frac{\gamma^2-1}{2}}\gamma^{-\frac{j+1}{2}},} & {(j\in \mathrm{odd})}, \\
    {\sqrt{\frac{\gamma^2-1}{2}}\gamma^{-\frac{N-j+1}{2}},} & {(j\in \mathrm{even})},
  \end{array}\right.
\end{align}
for $|\mathrm{KZM}_{\mathrm{P}}\rangle$ and
\begin{align}
  \chi_{j}=  \left\{\begin{array}{ll}
    {\sqrt{\frac{\gamma^2-1}{2}}\gamma^{-\frac{j+1}{2}},} & {(j\in \mathrm{odd}, j\neq N)}, \\
    {0,} & {(j=N)},\\
    {-\sqrt{\frac{\gamma^2-1}{2}}\gamma^{-\frac{N-j+1}{2}},} & {(j\in \mathrm{even})},
  \end{array}\right.
\end{align}
for $|\mathrm{KZM}_{\mathrm{A}}\rangle$ respectively. The solution is symmetric/antisymmetric on the two sides of the defect,
\be
  \psi_{N-j}&=&\psi_{j}, \\
  \chi_{N-j}&=&-\chi_{j}.
\ee
By taking the thermodynamic limit $N\rightarrow\infty$, one can find that the two KZM's are well localized near the defect as illustrated in Fig. 8.

%%%%%%%%%%% FIG 8  %%%%%%%%%%%%%%%%%%%%%%%%%%%%%%%
\begin{figure}[b]
\begin{center}
\includegraphics[width=3.5in,angle=0]{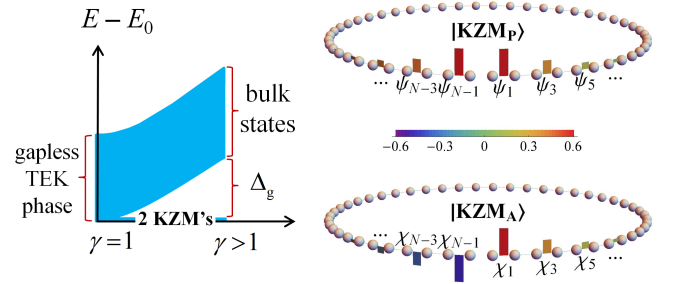}
\end{center}
\caption{(Color
  online) \emph{Left}: The mechanism for the localized KZM's emerging within the gapless TEK phase. An energy gap above the KZM's develops when the bulk band is split by modulating the parameter $\gamma=t_{N}/t$ to the range, $\gamma > 1$. \emph{Right}: Visualization of the two KZM's, $|\mathrm{KZM}_{\mathrm{P}}\rangle$ and $|\mathrm{KZM}_{\mathrm{A}}\rangle$. Here, we choose $N=51, U=1, t=0.1$, and $\gamma=2$ for demonstration. The coefficients, $\psi_{2}$, $\psi_{4}$, $\cdots$, $\psi_{N-2}$, $\psi_{N}$, $\chi_{2}$, $\chi_{4}$, $\cdots$, $\chi_{N-2}$, $\chi_{N}$, are not tagged because they are very small and neglectable.}%
\label{fig8}%
\end{figure}
%%%%%%%%%%% FIG 8  %%%%%%%%%%%%%%%%%%%%%%%%%%%%%%%

It is even more clear to see the picture of the KZM's by considering $\gamma\gg1$. In this limit, we get two simpler KZM's,
\be
  |\mathrm{KZM}_{\mathrm{P}}\rangle &\sim& (\psi_{1}, \psi_{N-1})= \frac{1}{\sqrt{2}}(1, 1),\\
  |\mathrm{KZM}_{\mathrm{A}}\rangle &\sim& (\chi_{1}, \chi_{N-1})= \frac{1}{\sqrt{2}}(1, -1).
\ee

\subsection{The gap beyond perturbative treatment}
In fact, we can calculate $H_{\mathrm{R}}(c)$ with defect at the symmetric point in a much more rigorous manner, because the QJWM still holds at this moment and the calculating scheme in Fig. 3 is still applicable. By the auxiliary Hamiltonians with defects, we have confirmed the emergence of the KZM's as depicted above. And by working out numerically the gap of systems as large as $N=1001$ and extrapolating to the the thermodynamic limit, we are able to correct the approximated gap formula Eq. (\ref{Delta_g}) to a rigorous one,
\be
  \Delta_{\mathrm{g}}=(\gamma+\frac{1}{\gamma}-2)~t~f(t/U),
\label{gapfit}
\ee
where the fitting function reads
\be
  f(t/U)\approx1-1.51726(t/U)+0.522561(t/U)^{2}.
\ee

\subsection{Influence of disorder}

The newly discovered KZM's is protected by the gap above them. However the gap is proportional to the hopping strength $t$, which should be small enough to validate the TEK phase. So we may ask whether a small disorder can destroy the gap and KZM's.

Let us consider the Hamiltonian at the symmetric point with disordered hopping strength,
\begin{align}
  H_{\mathrm{R}}(c)=\sum_{j=1}^{N}[t_{j}(-c_{j}^{\dag}c_{j+1} + c_{j}c_{j+1}+\mathrm{h.c.})\nonumber\\
 +U(2 n_{j}-1)(2 n_{j+1}-1)],  \label{Hdisorder}
\end{align}
where $t_{j} (j=1,2,...,N-1)$ is uniformly distributed in the interval $[t(1-\delta\tau/2),t(1+\delta\tau/2)]$ and $t_{N}$ takes values in the interval $[\gamma t(1-\delta\tau/2),\gamma t(1+\delta\tau/2)]$. Here, $\gamma=t_{N}/t$ still takes charge of the defect strength. We still consider the uniform interactions.

It is also easy to verify that the QJWM still holds and the calculating scheme in Fig. 3 is still applicable when the disorder is at presence. For disorder strength $\delta \tau$ ranging from 10\% to 40\%, we perform calculations on systems with  $N=501$ and over $1000$ random configurations. The results are illustrated in Fig. 9. From this result, we see the energy gap is stable for moderate disorder, which is important to protect the KZM's against disturbance.

%%%%%%%%%%% FIG 9  %%%%%%%%%%%%%%%%%%%%%%%%%%%%%%%
\begin{figure}[t]
\begin{center}
\includegraphics[width=3.2in,angle=0]{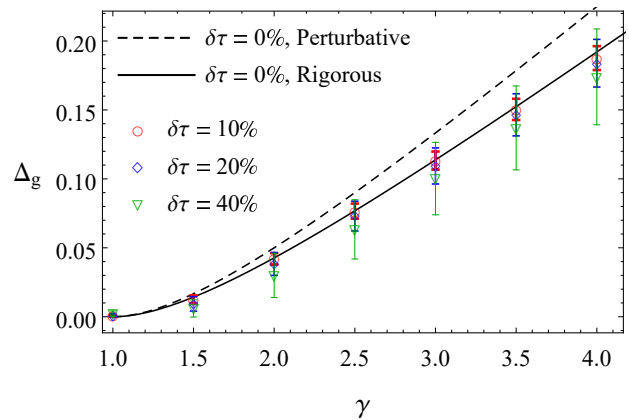}
\end{center}
\caption{(Color online) The energy gap above KZM's with disorder strength $\delta\tau$ ranging from 0\% to 40\%. Here we set $U=1$ and $t=0.1$. The dashed and solid lines show the results by perturbative theory and rigorous result for an infinite system without disorder ($\delta\tau=0\%$), i.e. Eqs. (\ref{Delta_g}) and (\ref{gapfit}) respectively. The data for disorder strength $\delta\tau$ ranging from 10\% to 40\% are obtained by performing calculations on systems with $N=501$ and over $1000$ random configurations.}
\label{fig9}%
\end{figure}
%%%%%%%%%%% FIG 9  %%%%%%%%%%%%%%%%%%%%%%%%%%%%%%%

\section{Summary and discussion}

In summary, a novel TEK phase and concomitant localized KZM's can be realized in the interacting Kitaev ring $H_{\mathrm{R}}(c)$ with GRF. Owing to the QJWM, this conclusion is also true for the other three rings, $H_{\mathrm{NS}}(c)$ and $H_{\mathrm{P/A}}(\sigma)$ with odd total number of lattice sites $N$.

Usually, the nontrivial topology is protected by a gap between the ground state and the bulk excitation spectra \cite{Shen}. Here, the gapless TEK phase and concomitant KZM's do not comply with the convention for noninteracting fermionic systems. Also, the simultaneous occurring of nondegeneracy of the ground state and quadratic gapless spectra in a clean ring seems odd in the ordinary field theory \cite{Franchini}.

We thank Yan He and Jian-Jun Dong for useful discussions. This work is supported by NSFC under Grants No. 11074177.

\appendix

\section{Exact solution of $H_{\mathrm{R}}(c)$}

According to the scheme depicted in Fig. 3, the solution of the Kitaev ring $H_{\mathrm{R}}(c)$ is resort to the four free fermion Hamiltonians,
\be
  H_{\mathrm{R}}^{\mathrm{(P)}}(f)&=&H(1,-t+U,t+U),\\
  H_{\mathrm{NS}}^{\mathrm{(P)}}(f)&=&H(-1,-t+U,t+U),\\
  H_{\mathrm{R}}^{\mathrm{(A)}}(f)&=&H(1,t+U,-t+U), \\
  H_{\mathrm{NS}}^{\mathrm{(A)}}(f)&=&H(-1,t+U,-t+U).
\ee
They can be cast into a same general expression,

\begin{widetext}

\begin{align}
  H(\xi, a, b) =\sum_{j=1}^{N-1}[a (f_{j}^{\dagger} f_{j+1}- f_{j} f_{j+1}^{\dagger})+b (f_{j}^{\dagger} f_{j+1}^{\dagger}- f_{j} f_{j+1})] +\xi [a (f_{N}^{\dagger} f_{1}- f_{N} f_{1}^{\dagger})+b (f_{N}^{\dagger} f_{1}^{\dagger}- f_{N} f_{1})]
\end{align}
After Fourier transformation, we get
\be
  H(\mathrm{s},a,b)=\sum_{q\in Q_{\mathrm{s}}, q\neq q_{\mathrm{s}}}
                  [2a\cos q(f_{q}^{\dag}f_{q}-f_{-q}f_{-q}^{\dag})
            -2\mathrm{i}b\sin q(f_{q}^{\dag}f_{-q}^{\dag}-\mathrm{h.c.})]+a\cos q_{\mathrm{s}}(2 f_{q_{\mathrm{s}}}^{\dag}f_{q_{\mathrm{s}}}-1),
\ee
\end{widetext}

where $\mathrm{s}$ refers to R (or $\xi=1$) or NS (or $\xi=-1$), $q_{\mathrm{R}}=0, q_{\mathrm{NS}}=\pi$, and
\be
  &Q_{\mathrm{R}}=\{-\frac{N-1}{N}\pi,\ldots,-\frac{2}{N}\pi,0,\frac{2}{N}\pi,\ldots,\frac{N-1}{N}\pi\}, \\
  &Q_{\mathrm{NS}}=\{-\frac{N-2}{N}\pi,\ldots,-\frac{1}{N}\pi,\frac{1}{N}\pi,\ldots,\frac{N-2}{N}\pi,\pi\}.
\ee
By Bogoliubov transformation,
\be
  \eta_{q}=&u_{q}f_{q}-\operatorname{i}v_{q}f_{-q}^{\dagger},~~(q\neq 0,\pi),
\ee
where
\be
  &u_{q}^{2}=\frac{1}{2}\left(  1+\frac{\epsilon(q)}{\omega(q)}\right)
    ,v_{q}^{2}=\frac{1}{2}\left(  1-\frac{\epsilon(q)}{\omega(q)}\right)
    ,\nonumber\\
  &2u_{q}v_{q}  =\frac{D(q)}{\omega(q)},\\
  &\epsilon(q)=a\cos{q},D(q)=b\sin{q},\\
  &\omega(q)=\sqrt{\epsilon(q)^{2}+D(q)^{2}},
\ee
we can diagonalize the fermionic Hamiltonian by arriving at
\begin{align}
  H(\mathrm{s},a,b)=\epsilon(q_{\mathrm{s}})( 2 f_{q_{\mathrm{s}}}^{\dagger} f_{q_{\mathrm{s}}}-1)+\sum_{q\in Q_{\mathrm{s}}, q\neq q_{\mathrm{s}}}\omega(q)\left(  2\eta_{q}^{\dagger}\eta_{q}-1\right)  .
\label{H(o)}%
\end{align}
There are four fermion vacua in BCS-type wave functions,
\begin{align}
|\phi_{\mathrm{R/NS}}^{\mathrm{(P)/(A)}}\rangle &  =\prod_{\substack{q\in Q_{\mathrm{R}/\mathrm{NS}}, \\(0<q<\pi)}}\left(
u_{q}+\operatorname{i}v_{q} f_{q}^{\dag} f_{-q}^{\dag}\right)  |0\rangle
\end{align}
for $H_{\mathrm{R/NS}}^{\mathrm{(P)/(A)}}$ respectively. The two channels, $H_{\mathrm{P/A}}(\sigma)$, provide two quasiparticle spectra,
\be
  \omega_{\mathrm{P}}(q)=\sqrt{U^2+t^2-2 U t\cos(2q)}, \\
  \omega_{\mathrm{A}}(q)=\sqrt{U^2+t^2+2 U t\cos(2q)}.
\ee
The solution of $H_{\mathrm{R}}(c)$ is obtained by filtering out redundant DOF by applying the scheme in Fig. 3 backwardly.

Of all the energy levels, we focus on the $2N$ extended-kink (EK) states. Half of them are of odd parity and form the spectrum,
\be
  E_{\mathrm{P}}(q)=2\omega_{\mathrm{P}}(q)+\mathscr{E}_{0}-(U-t),
\ee
while another half of them are of even parity and form the spectrum
\be
  E_{\mathrm{A}}(q)=2\omega_{\mathrm{A}}(q)+\mathscr{E}_{0}+|U+t|-2U,
\ee
where
\begin{equation}
  \mathscr{E}_{0}=(U-t)-\sum_{q\in Q_{\mathrm{R}}}\omega_{\mathrm{P}}(q). \label{constantE0}
\end{equation}

We should notice that the eigenstates of $H_{\mathrm{R/NS}}^{\mathrm{(P)/(A)}}$ are ones of the parity operator $\mathcal{P}_{y} = \exp(i \pi M_{y})$ with $M_{y}=\sum_{j=1}^{N} f_{j}^{\dagger}f_{j}$, not the ones of parity operator $\mathcal{P}_{z}$. So a further projections by projectors $P_{y}^{\pm}=\frac{1}{2}(\hat{1}\pm\mathcal{P}_{y})$ are needed. The validity of this route is ensued by the property ($f$-fermion particle-hole transformation),
\begin{align}
  &\mathscr{C}^{\dag}_{f} H_{\mathrm{R}}^{\mathrm{(P)/(A)}}(f) \mathscr{C}_{f} = H_{\mathrm{NS}}^{\mathrm{(P)/(A)}}(f),\\
  &\mathscr{C}_{f}=\prod_{j=1}^{N}[f_{j}^{\dag}+(-1)^{j}f_{j}].
\end{align}
This property means the energy levels of $H_{\mathrm{R}}^{\mathrm{(P)/(A)}}(f)$ and $H_{\mathrm{NS}}^{\mathrm{(P)/(A)}}(f)$ are the same. For example, in $H_{\mathrm{P}}(\sigma)$ channel, the ground states of $H_{\mathrm{R}}^{\mathrm{(P)}}(f)$ and $H_{\mathrm{NS}}^{\mathrm{(P)}}(f)$ are $f_{0}^{\dag}|\phi_{\mathrm{R}}^{\mathrm{(P)}}\rangle$ and $|\phi_{\mathrm{NS}}^{\mathrm{(P)}}\rangle$ respectively, i.e., they have the same energy value $E_0$,
\be
  H_{\mathrm{R}}^{\mathrm{(P)}}(f) f_{0}^{\dag}|\phi_{\mathrm{R}}^{\mathrm{(P)}}\rangle &=& E_0 f_{0}^{\dag}|\phi_{\mathrm{R}}^{\mathrm{(P)}}\rangle,\\
  H_{\mathrm{NS}}^{\mathrm{(P)}}(f) |\phi_{\mathrm{NS}}^{\mathrm{(P)}}\rangle &=& E_0 |\phi_{\mathrm{NS}}^{\mathrm{(P)}}\rangle.
\ee
But their parities in $y$ direction are opposite,
\be
  \mathscr{P}_{y} f_{0}^{\dag}|\phi_{\mathrm{R}}^{\mathrm{(P)}}\rangle &=& -f_{0}^{\dag}|\phi_{\mathrm{R}}^{\mathrm{(P)}}\rangle,\\
  \mathscr{P}_{y} |\phi_{\mathrm{NS}}^{\mathrm{(P)}}\rangle &=& + |\phi_{\mathrm{NS}}^{\mathrm{(P)}}\rangle.
\ee
Because $\{\mathscr{P}_{y}, \mathscr{P}_{z}\}=0$ and $\mathscr{P}_{y}^2=\mathscr{P}_{z}^2=1$, we can find that
\begin{align}
  \mathscr{P}_{z} (f_{0}^{\dag}|\phi_{\mathrm{R}}^{\mathrm{(P)}}\rangle+|\phi_{\mathrm{NS}}^{\mathrm{(P)}}\rangle) = + (f_{0}^{\dag}|\phi_{\mathrm{R}}^{\mathrm{(P)}}\rangle+|\phi_{\mathrm{NS}}^{\mathrm{(P)}}\rangle),\\
  \mathscr{P}_{z} (f_{0}^{\dag}|\phi_{\mathrm{R}}^{\mathrm{(P)}}\rangle-|\phi_{\mathrm{NS}}^{\mathrm{(P)}}\rangle) = - (f_{0}^{\dag}|\phi_{\mathrm{R}}^{\mathrm{(P)}}\rangle-|\phi_{\mathrm{NS}}^{\mathrm{(P)}}\rangle).
\end{align}
For finite $N$ and in the range $t>0$ and $-\infty<U<\infty$, the non-degenerate ground state is of odd parity ($\mathscr{P}_{z}=-1$) and reads
\be
  |E_0\rangle =|E_{\mathrm{P}}(0)\rangle= \frac{1}{\sqrt{2}}(f_{0}^{\dag}|\phi_{\mathrm{R}}^{\mathrm{(P)}}\rangle-|\phi_{\mathrm{NS}}^{\mathrm{(P)}}\rangle).
\ee
The upper EK states can be treated in the subspaces of the same energy value likewise.

\section{Correlation function of the ground state}

\subsection{Toeplitz determinant representation}

The correlation function is defined as
\be
  C_{r,N}&=&\langle E_{0}|(2 n_{j}-1)(2 n_{j+r}-1)|E_{0}\rangle\nonumber\\
     &=&\langle E_{0}|\sigma_{j}^{z}\sigma_{j+r}^{z}|E_{0}\rangle.\label{crNB}
\ee
Due to the parity, we have $  \langle\phi_{\mathrm{R}}^{(P)}|f_{0}\sigma_{j}^{z}\sigma_{j+r}^{z}|\phi_{\mathrm{NS}}^{(P)}\rangle=0$. By introducing the Majorana fermions,
\be
  A_{l}=f_{l}^{\dagger}+f_{l}, B_{l}=f_{l}^{\dagger}-f_{l},
\ee
we can expand the product of two spin operators as
\begin{equation}
\sigma_{j}^{z}\sigma_{j+r}^{z}=B_{j}A_{j+1}B_{j+1}\ldots A_{j+r-1}B_{j+r-1}A_{j+r}.
\end{equation}
Then by using Wick's theorem basing on the relations,
\begin{align}
  &\langle f_{0}f_{0}^{\dag}\rangle=1, \nonumber\\
  &\langle A_{j}f_{0}^{\dag}\rangle=-\langle B_{j}f_{0}^{\dag}\rangle=\frac{1}{\sqrt{N}}, \nonumber\\
  &\langle A_{i}A_{j}\rangle=-\langle B_{i}B_{j}\rangle=\delta_{i,j}, \nonumber\\
  &\langle B_{i}A_{i+r}\rangle=-\langle A_{i+r}B_{i}\rangle=\mathscr{D}^{\mathrm{(s)}}_{r+1},
\end{align}
we get the Toeplitz determinant representation of the correlation function,

\be
  C_{r,N}=\frac{1}{2}\left(\det[\mathscr{D}^{\mathrm{(R)}}_{l-m}+\frac{2}{N}]
  +\det[\mathscr{D}^{\mathrm{(NS)}}_{l-m}]\right), \label{toe}
\ee
where
\begin{align}
  &\det[\mathscr{D}^{\mathrm{(R)}}_{l-m}+\frac{2}{N}]  \nonumber\\
  &=\left\vert
    \begin{array}[c]{cccc}%
      \mathscr{D}^{\mathrm{(R)}}_{0}+\frac{2}{N} & \mathscr{D}^{\mathrm{(R)}}_{-1}+\frac{2}{N} & \cdots & \mathscr{D}^{\mathrm{(R)}}_{1-r}+\frac{2}{N}\\
      \mathscr{D}^{\mathrm{(R)}}_{1}+\frac{2}{N} & \mathscr{D}^{\mathrm{(R)}}_{0}+\frac{2}{N} & \cdots & \mathscr{D}^{\mathrm{(R)}}_{2-r}+\frac{2}{N}\\
      \cdots & \cdots & \cdots & \cdots\\
      \mathscr{D}^{\mathrm{(R)}}_{r-1}+\frac{2}{N} & \mathscr{D}^{\mathrm{(R)}}_{r-2}+\frac{2}{N} & \cdots & \mathscr{D}^{\mathrm{(R)}}_{0}+\frac{2}{N}%
    \end{array}
  \right\vert,
\end{align}
\begin{align}
  \det[\mathscr{D}^{\mathrm{(NS)}}_{l-m}]=\left\vert
    \begin{array}[c]{cccc}%
      \mathscr{D}^{\mathrm{(NS)}}_{0} & \mathscr{D}^{\mathrm{(NS)}}_{-1} & \cdots & \mathscr{D}^{\mathrm{(NS)}}_{1-r}\\
      \mathscr{D}^{\mathrm{(NS)}}_{1} & \mathscr{D}^{\mathrm{(NS)}}_{0} & \cdots & \mathscr{D}^{\mathrm{(NS)}}_{2-r}\\
      \cdots & \cdots & \cdots & \cdots\\
      \mathscr{D}^{\mathrm{(NS)}}_{r-1} & \mathscr{D}^{\mathrm{(NS)}}_{r-2} & \cdots & \mathscr{D}^{\mathrm{(NS)}}_{0}%
    \end{array}
  \right\vert.
\end{align}
and
\begin{align}
\mathscr{D}^{\mathrm{(s)}}_{n}&=-\frac{e^{-i q_{\mathrm{s}}}}{N}e^{-i q_{\mathrm{s}}n}+
\frac{1}{N}\sum_{q\in Q_{\mathrm{s}}, q\neq q_{\mathrm{s}}}e^{-i q n}D(e^{i q}),\nonumber\\
&D(e^{i q})=-\frac{U-t e^{2i q}}{\sqrt{(U-t e^{2iq})(U-t e^{-2i q})}}.
\end{align}

\subsection{Evaluation of the Toeplitz determinants}

In the thermodynamic limit, $N\rightarrow\infty$, we can work out the two Toeplitz determinants in Eq. (\ref{toe}) by applying the \emph{Generalized Szeg\"{o}'s Theorem} \cite{DongJSM, DongPRE} and obtain

\begin{align}
  \det[\mathscr{D}^{\mathrm{(R)}}_{l-m}+\frac{2}{N}] = \det[\mathscr{D}^{\mathrm{(NS)}}_{l-m}] =\Delta_{r}(1+\frac{x r}{N D(e^{iq})}),
\end{align}
where
\begin{align}
  &\Delta_{r}= u^{r}\exp(\sum_{n=1}^{\infty}nd_{-n}d_{n}),\\
  &u=\exp\{\int_{-\pi}^{\pi}\frac{dq}{2\pi}\ln{D(e^{iq})}\},\\
  &d_{n}=\int_{-\pi}^{\pi}\frac{dq}{2\pi}e^{-iqn}\ln{D(e^{iq})}.
\end{align}

Let us see the TEK phase ($U>t$) first. Noticing that
\begin{align}
  D(e^{iq})&=-\frac{U-te^{2iq}}{\sqrt{(U-te^{2iq})(U-te^{-2iq})}}
  \nonumber\\&=-\sqrt{\frac{1-\lambda e^{2iq}}{1-\lambda e^{-2iq}}},
\end{align}
where
\begin{align}
  &\lambda=t/U,\\
  &\ln \left(1-\lambda e^{i q}\right)=-\sum_{n=1}^{\infty} \frac{1}{n}\left(\lambda e^{i q}\right)^{n},
\end{align}
we can deduce in the following way,
\begin{align}
  &\ln D\left(e^{i q}\right) =i \pi-\frac{1}{2} \ln \left(1-\lambda e^{-2i q}\right)+\frac{1}{2} \ln \left(1-\lambda e^{2iq}\right) \nonumber\\ &=i \pi+\frac{1}{2} \sum_{n=1}^{\infty} \frac{\lambda^{n}}{n} e^{-2i q n}-\frac{1}{2} \sum_{n=1}^{\infty} \frac{\lambda^{n}}{n} e^{2 i q n},
\end{align}
so as to get
\begin{equation}
\mu=\exp \left[\int_{-\pi}^{\pi} \frac{d q}{2 \pi} \ln D\left(e^{i q}\right)\right]=-1
\end{equation}
and ($d_{2k+1}=0$)
\begin{align}
  &d_{2k}=-\frac{\lambda^{k}}{2k}, d_{-2k}=\frac{\lambda^{k}}{2k}, \\
  &\sum_{n=1}^{\infty} n d_{n} d_{-n}=-\sum_{k=1}^{\infty} \frac{1}{2k} \lambda^{2 k}=\frac{1}{2} \ln \left(1-\lambda^{2}\right).
\end{align}
Thus the Toeplitz determinant is worked out as
\begin{equation}
\Delta_{r}(1+\frac{x r}{N D(e^{iq})})=(-1)^{r}\sqrt{1-t^{2}/U^{2}}(1-\frac{2r}{N}),
\end{equation}
which means the correlation function of the TEK phase reads
\begin{equation}
C(r,\alpha)=(-1)^{r}\sqrt{1-t^{2}/U^{2}}(1-\frac{2r}{N}).
\end{equation}

For the CAT ($U<-t$) and TSC ($|U|<t$) phases, we can apply similar analysis. Eventually, we obtain the correlation function in the three phases,
\begin{align}
  C(r,\alpha)&\equiv\lim_{N\rightarrow\infty}C_{r,N}\nonumber\\
&\approx \left\{\begin{array}{ll}
    {\sqrt{1-t^{2}/U^{2}},} & {(U<-t)}, \\
    {0,} & {(|U|<t)}, \\
    {(-1)^{r}\sqrt{1-t^{2}/U^{2}}(1-2\alpha),} & {(U>t)},
  \end{array}\right. \label{Calpha}
\end{align}
where $\alpha=\frac{r}{N}$.


\begin{thebibliography}{99}
\bibitem {explain} These bound states should not be confused with the Majorana bound states, or Majorana zero modes.
\end{thebibliography}
\end{document}